\documentclass[5p]{elsarticle}

\usepackage{amssymb}
\usepackage{latexsym}

\usepackage{bm}
\usepackage{color}
\usepackage{hyperref}

\begin{document}


\begin{frontmatter}

\title{Newman-Ziff algorithm for the bootstrap percolation: application to the Archimedean lattices}

\author[1]{Jeong-Ok Choi}
\author[2]{Unjong Yu\corref{cor1}}
\cortext[cor1]{Corresponding author: 
  Tel.: +82-62-715-3629; }
\ead{uyu@gist.ac.kr}

\address[1]{Division of Liberal Arts and Sciences, 
            Gwangju Institute of Science and Technology,
            Gwangju 61005, South Korea}
\address[2]{Department of Physics and Photon Science,
            Gwangju Institute of Science and Technology, 
            Gwangju 61005, South Korea}


\begin{abstract}
We propose very efficient algorithms for the bootstrap percolation and the diffusion percolation models
by extending the Newman-Ziff algorithm of the classical percolation
[M.~E.~J.~Newman and R.~M.~Ziff, Phys. Rev. Lett. 85 (2000) 4104].
Using these algorithms and the finite-size-scaling, 
we calculated with high precision the percolation threshold
and critical exponents in the eleven two-dimensional Archimedean lattices.
We present the condition for the continuous percolation transition 
in the bootstrap percolation and the diffusion percolation,
and show that they have the same critical exponents as the classical percolation within error bars in two dimensions.
We conclude that the bootstrap percolation and the diffusion percolation
almost certainly belong to the same universality class as the classical percolation.
\end{abstract}


\end{frontmatter}


\section{Introduction}
The bootstrap percolation (BP) model \cite{Chalupa79} has attracted continuous attention
for its various applications such as disordered dilute magnetic systems 
\cite{Chalupa79,Sabhapandit02}, neuronal activity \cite{Soriano08}, 
jamming transition \cite{Gregorio05}, and diffusion of innovations \cite{Helbing12}.
The BP process operates as follows: 
(i) Each site is occupied with the probability $p$ and empty otherwise; 
(ii) Occupied sites that have less than $m$ occupied neighbors become empty, 
and the process is repeated until all the occupied sites have at least $m$ occupied neighbors.
The diffusion percolation (DP) model \cite{Adler88} is closely-related to the BP,
and sometimes it is also called the BP \cite{Baxter10,Gravner12,Gao15}.
In the DP, process (i) is the same as the BP, but process (ii) is different: 
empty sites that have at least $k$ occupied neighbors become occupied recursively,
until all the empty sites have less than $k$ occupied neighbors.

The BP and DP have been studied on lattices 
\cite{Kogut81,Branco84,Khan85,Branco86,Adler88,Adler90,Adler91,Chaves95,Medeiros97,Branco99,Gravner12},
trees \cite{Chalupa79,Fontes08,Bollobas14}, 
and complex networks \cite{Balogh07,Baxter10,Wu14,Gao15}.
A few facts are known about the BP and DP.
Clearly, the BP with $m=0$ and DP with $k>\Delta_{\mathrm{max}}$ are the same as the classical percolation (CP) model,
where $\Delta_{\mathrm{max}}$ is the maximum value of degree.
For a given graph, the percolation threshold of the DP ($p_{\mathrm{\scriptscriptstyle DP}}^k$) is not larger than that of the CP ($p_{\mathrm{\scriptscriptstyle CP}}^{}$), 
and that of the BP ($p_{\mathrm{\scriptscriptstyle BP}}^m$) is not less than that of the CP.
The BP with $m=1$ or $m=2$ has the same percolation threshold as the CP. 
As for $\Delta$-regular lattices, if $m+k=\Delta+1$, then there is a close relationship between the BP and DP:
(1) The sum of percolation thresholds of the BP and its corresponding DP is the same or larger than 1
($p_{\mathrm{\scriptscriptstyle BP}}^m+p_{\mathrm{\scriptscriptstyle DP}}^{\Delta+1-m} \geq 1$) \cite{Adler88}
(The sum is 1 for self-matching lattices such as the triangular lattice \cite{Sykes64}.);
(2) There are lattice-dependent parameters $m_c$ and $k_c$, 
where the BP of $m$ and DP of $k$ have
first-order percolation transitions with the percolation thresholds 
$p_{\mathrm{\scriptscriptstyle BP}}^m=1$ and $p_{\mathrm{\scriptscriptstyle DP}}^k=0$, respectively,
if and only if $m>m_c$ and $k<k_c$; the two parameters satisfy $m_c + k_c = \Delta+1$.
For the square and triangular lattices, $m_c = \Delta/2$ \cite{Kogut81,Enter87,Adler88}.

About the critical exponents and the universality class 
of continuous percolation transitions of the BP and DP,
there have been long debates. 
There are several critical exponents for the percolation,
and in fact any two of them determine all of the rest exponents through scaling relations \cite{Stauffer94}.
For the CP, they are universal and depend only on the spatial dimension. 
While the BP of $m=1$ is known to have the same critical exponents as the CP \cite{Adler90}, 
it is not clear for the DP and the BP of $m\geq 2$.
Kogut and Leath argued that $\beta$ depends on $m$ for the BP
from Monte-Carlo simulations on the square, triangular, 
and cubic lattices \cite{Kogut81},
and renormalization group studies confirmed it \cite{Branco84,Branco86}.
Adler conjectured that $\nu$ is universal but $\beta$ is non-universal in the BP and DP \cite{Adler88,Adler90,Adler91}.
To the contrary, other studies, which include most recent simulations,
insist that both $\nu$ and $\beta$ are universal \cite{Khan85,Chaves95,Medeiros97,Branco99}.
They argued that non-universality of previous works can be
from the small size of clusters used in the simulations.
However, we judge that the universality class 
of the BP and DP is not definitely clear yet.
In Ref.~\cite{Medeiros97}, for example, the Fisher exponent $\tau$ for
the DP with $k=4$ was calculated on the triangular lattice 
to be $\tau = 2.03\pm 0.04$, which is consistent with
$\tau=187/91\approx 2.055$ for the two-dimensional CP. 
At first glance, it looks like a good evidence
to support the conclusion of the same universality;
however, $\beta$ obtained by the scaling relation 
$\beta/\nu = 2(\tau-2)/(\tau-1)$ in two dimensions is 
$\beta/\nu = 0.06\pm 0.08$,
which is consistent with $\beta/\nu = 5/48\approx 0.104$ of the CP 
but the uncertainty is too large to make any conclusion.
In addition, the BP and DP have been studied only in four kinds
of lattices (the square, triangular, honeycomb, and cubic lattices). 

In this paper, we introduce efficient algorithms for the BP and DP models
and present much more precise results on eleven two-dimensional Archimedean lattices.
We calculated percolation threshold and critical exponents ($\nu$ and $\beta$)
to present positive evidences for that the BP and DP belong to the same universality class as the CP in two dimensions.

\begin{figure*}[!tb]
\centering
\includegraphics[width=15cm]{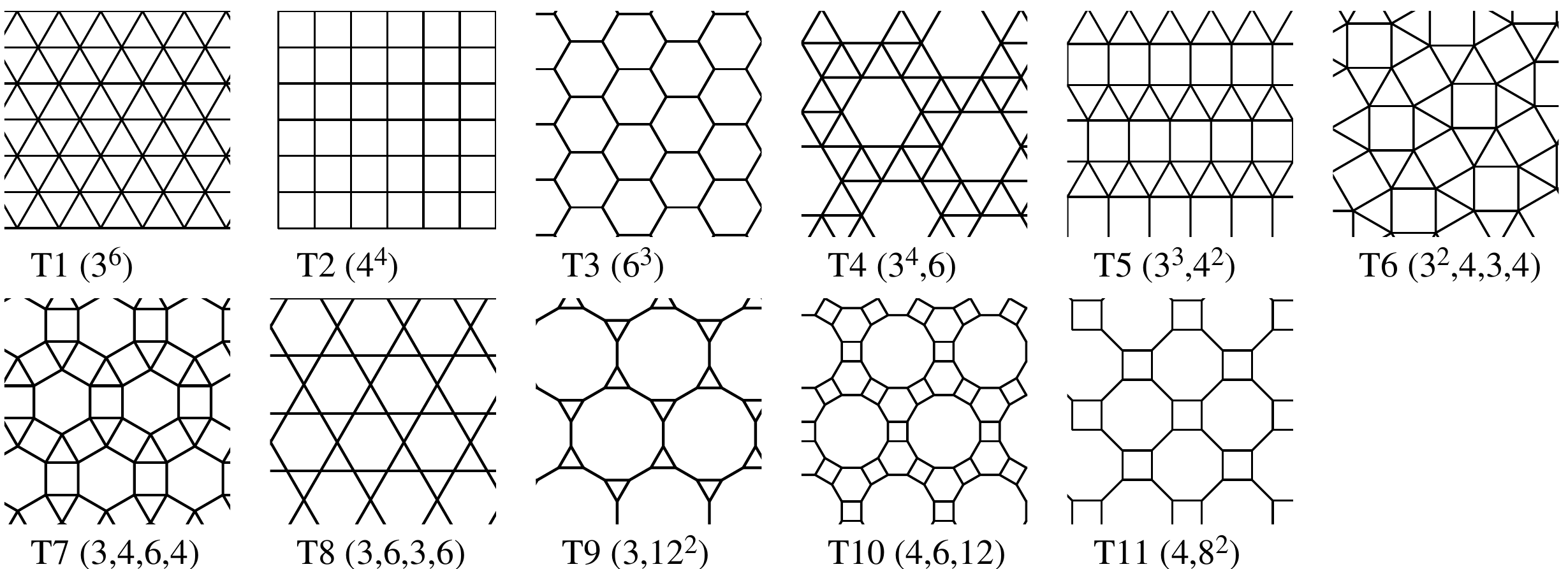}
\caption{The eleven Archimedean lattices. The numbers in parentheses represent the sequence of regular polygons around each vertex.}
\label{Arch}
\end{figure*}

\section{Methods}

The classical site percolation model can be simulated simply 
by the following two steps:
(i) Fill each site with independent probability $p$ and 
leave it empty with probability $1-p$; (ii) Identify all the
connected clusters to check whether a percolating cluster exists.
The simulation should be repeated many times to make a statistical average for a given $p$.
This algorithm is simple but inefficient. In order to get results with different
probability $p'$, the whole simulation should be done again. 
It is more difficult to calculate the derivative of a quantity with respect to the probability $p$,
because numerical differentiation inevitably gives large error \cite{Klein06}.
A more efficient algorithm was proposed by Newman and Ziff \cite{Newman00,Newman01},
and it has become a standard method in classical percolation studies.
The Newman-Ziff algorithm consists of four steps:
(i) Initially, all sites are empty;
(ii) Choose an empty site randomly and fill it;
(iii) Update the information of connected clusters to check whether percolation occurs;
(iv) Repeat steps (ii) and (iii) until all the sites are occupied.
An efficient algorithm to update the information of connected clusters
(tree-based union/find algorithms) is 
also presented in Ref.~\cite{Newman01}.
An average $\langle Q(n) \rangle$ is obtained by repeating the whole steps,
where $Q(n)$ is any quantity (e.g., size of the largest cluster)
for a fixed number of occupied sites $n$.
In one run of the Newman-Ziff algorithm, 
$Q(n_1)$ is correlated with $Q(n_2)$ inevitably, 
but values of $Q(n)$ of different runs are absolutely independent 
and so the statistical averaging of $\langle Q(n) \rangle$ has no problem.
A value $\langle Q(p) \rangle$ for a fixed occupation probability $p$
can be obtained by the transformation of
\begin{eqnarray}
\langle Q(p) \rangle = \sum_{n=0}^{N} \frac{N!}{n! (N-n)!}
                       p^n (1-p)^{N-n}
                       \; \langle Q(n) \rangle , \label{caninical_transf}
\end{eqnarray}
where $N$ is the total number of sites. 
Therefore, once $\langle Q(n) \rangle$ is obtained, 
$\langle Q(p) \rangle$ can be calculated for all values of $p$.
Another advantage of the Newman-Ziff algorithm is that the derivative
can be obtained through
\begin{eqnarray}
\frac{d\langle Q(p) \rangle}{dp} = \frac{d}{dp} \left[\sum_{n=0}^{N} \frac{N!}{n! (N-n)!}
                                    p^n (1-p)^{N-n}
                                    \; \langle Q(n) \rangle \right] \\
= \sum_{n=0}^{N} \frac{N!}{n! (N-n)!} \; p^{n-1} (1-p)^{N-n-1}(n-Np) \; 
    \langle Q(n) \rangle 
\end{eqnarray}
without numerical differentiation \cite{Martins03}.

However, the Newman-Ziff algorithm cannot be used directly in the BP or DP models,
because filling of each site depends on the local environment.
As for the DP, the Newman-Ziff algorithm can be modified as follows. \\ \\
(1) Initially, all sites are empty.\\
(2) Make an array of all the sites in random order.\\
(3) Get one site by the array.
    If the site is empty, fill it and other sites
    that have at least $k$ occupied neighbors, recursively.
    If the site is already occupied, do nothing. \\
(4) Update the information about connected clusters 
    to check whether percolation occurs. \\
(5) Repeat steps (3) and (4) until all the sites are occupied. \\ \\
The whole steps are repeated to make an average quantity $\langle Q(n) \rangle$,
and the transformation of Eq.~(\ref{caninical_transf}) gives $\langle Q(p) \rangle$.
This algorithm is equivalent to the original DP with fixed $p$
because the final state does not depend on the sequence of the filling process
once an initial occupation is determined.
The BP model can be simulated by the same way as DP:
Initially all sites are filled and sites are emptied one-by-one in random order. 
This algorithm is called the avalanching bootstrap percolation 
of the second kind (ABP2) \cite{Farrow05}.
However, the algorithm is inefficient because it is difficult to identify 
and update the cluster information
during simulation. Therefore, we propose a more efficient algorithm 
for the BP model by introducing
\textit{preoccupied} state in addition to empty and occupied states.
When a site is chosen in the Newman-Ziff algorithm, 
it is occupied only when there are at least $m$ occupied neighbors; 
otherwise, it is assigned to be preoccupied. 
Preoccupied state means that the site will be occupied after the condition is satisfied. 
The algorithm can be presented as follows.\\ \\
(1) Initially, all sites are empty.\\
(2) Make an array of all the sites in random order.\\
(3-a) Get one site by the array.
    Set the site into the preoccupied state. \\
(3-b) Identify the connected cluster of preoccupied sites that includes the site.
    (To accelerate the simulation, preoccupied sites
     with less than $m$ occupied or preoccupied neighbors can be excluded from the cluster,
     because they are impossible to be filled.) \\
(3-c) Fill tentatively all the sites of the cluster. \\
(3-d) Within the cluster, set sites that have less than $m$ occupied neighbors
    into the preoccupied state again, recursively,
    until all the occupied sites have at least $m$ occupied neighbors. \\
(4) Update the information about connected clusters to check whether percolation occurs. \\
(5) Repeat steps (3) and (4) until all the sites are occupied. \\ \\
All steps of this algorithm are identical with the Newman-Ziff algorithm of the CP except for step (3), 
and the efficient routine that updates the information of connected clusters
can be used without modification.
Although we focus on two-dimensional lattices in this paper, both of the algorithms for the DP and BP can also be applied to any dimensional systems and to complex networks.

The CPU time needed in these algorithms is proportional to the lattice size by $T_{\mathrm{cpu}} \sim N^{\lambda}$ with $\lambda \approx 1.2$. CPU time for the DP, CP, and BP ($m<3$) are of the same order of magnitude for the same value of $N$; 
in the case of the BP with $m=3$, CPU time requirement is about 10 times more than that of the CP. 
Three-dimensional lattices show the same behavior, but more calculations would be needed than two dimensions because the number of sites increases as $N \sim L^3$, where $L$ is linear size. 
For a lattice of $N=10^6$, it takes about one second of CPU time for one sweep except for the BP of $m=3$ by Intel(R) Xeon(R) CPU of 2.2 GHz. 
Note that the running time can be easily reduced by parallelizing the algorithm.
In the case of the traditional brute force approach, which calculates $\langle Q(p) \rangle$ directly, the CPU time requirement depends on the number of occupied sites. 
On average, it takes about half of the Newman-Ziff algorithm because some part of the lattice is not filled at all.
However, it gives only $\langle Q(p) \rangle$ at a specific value of $p$, and another independent calculation is needed to get $\langle Q(p') \rangle$ for $p' \neq p$. 
In addition, it is practically impossible to get its derivative $d\langle Q(p) \rangle/dp$ with high precision. 
Therefore, our new algorithms are much more efficient except when only $\langle Q(p) \rangle$ is to be calculated at a known specific $p$.

Using these algorithms, we calculated the strength of the largest cluster
(the probability that a site belongs to the largest cluster; $P_{\infty}$),
average cluster size excluding the largest one ($M'_{1}$), percolation probability in 
any direction ($P_{w1}$) and in both directions ($P_{w2}$), and proportion of
occupied sites ($P_{o}$) as a function of $p$ for the CP, BP, and DP.

In this work, we consider eleven Archimedean lattices, which are vertex-transitive graphs
made in two dimensions by edge-to-edge tiling of regular polygons
whose vertices are surrounded by the same sequence of polygons.
There are only eleven Archimedean lattices \cite{Grunbaum87}
and they are typically used for systematic studies \cite{Suding99,Richter04,Yu17}.
They are shown in Fig.~\ref{Arch}.
The periodic boundary condition is used and the percolation is defined by the existence of a cluster that wraps all the way around the lattice.
The number of lattice sites studied in this work is from $N=1296$ to $N=2\,560\,000$.

\begin{figure*}[!b]
\centering
\includegraphics[width=16cm]{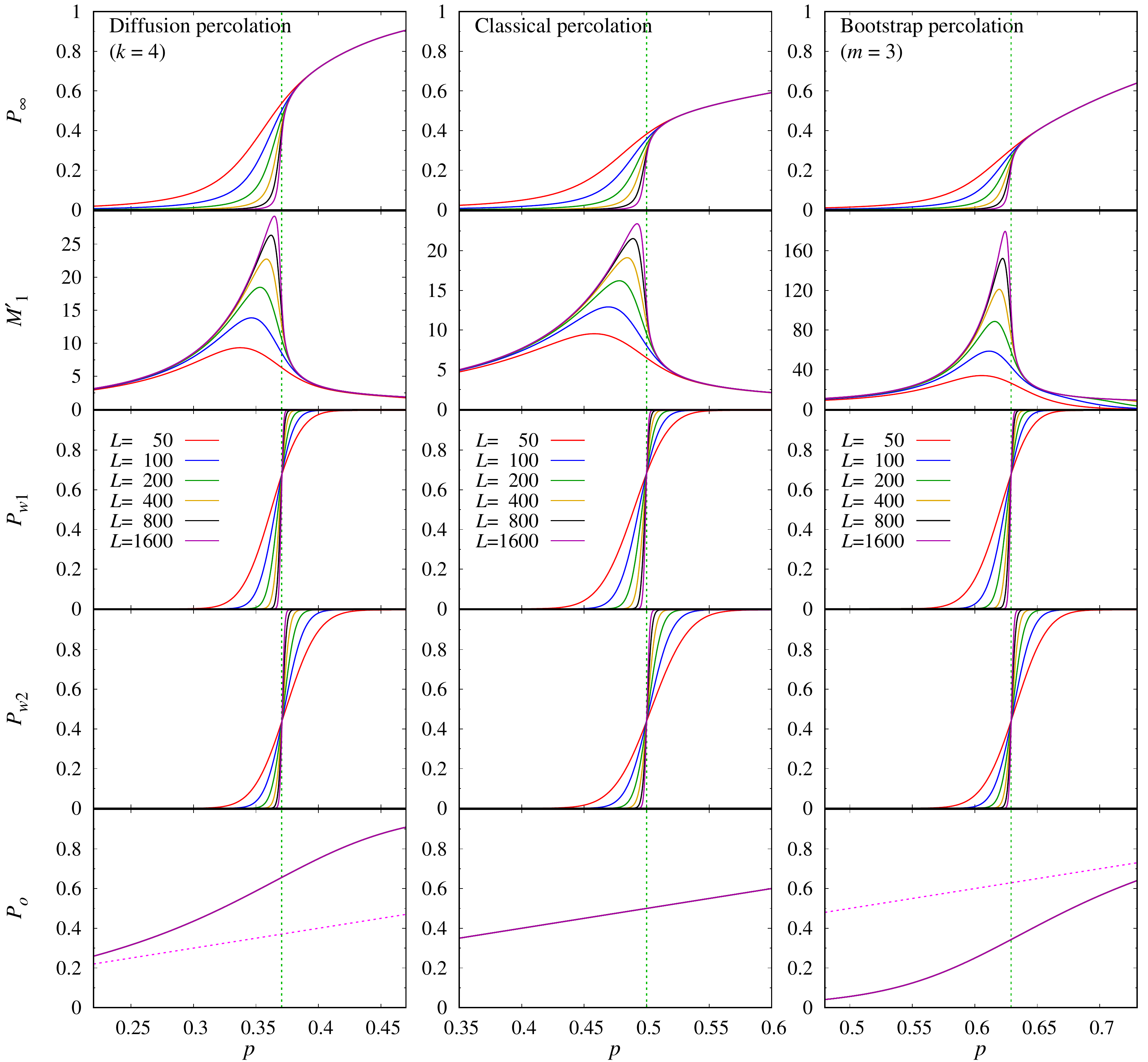}
\caption{Strength of the largest cluster ($P_{\infty}$),
average cluster size excluding the largest one ($M'_{1}$), 
percolation probability in any direction ($P_{w1}$) and in both directions ($P_{w2}$), and proportion of
occupied sites ($P_{o}$) as a function of initial filling probability ($p$) 
for the DP ($k=4$), CP, and BP ($m=3$)
in the triangular lattice (T1) for various linear size $L$. 
Vertical dotted lines indicate our estimates of percolation thresholds
($p_{\mathrm{\scriptscriptstyle DP}}^k$, $p_{\mathrm{\scriptscriptstyle CP}}^{}$, 
and $p_{\mathrm{\scriptscriptstyle BP}}^m$).
Diagonal dotted straight lines in the lowest row represent $P_o = p$.}
\label{P_vs_p}
\end{figure*}

The percolation thresholds ($p_{\mathrm{\scriptscriptstyle DP}}^k$ and $p_{\mathrm{\scriptscriptstyle BP}}^m$) 
of infinite lattices
are determined by the finite-size-scaling.
The percolation threshold estimate of a finite lattice is
determined by the probabilities of initial filling ($p$) 
that give the maximums of physical quantities that show critical behavior or their derivatives.
The percolation threshold can also be found by the crossing points 
of percolation probabilities of different lattice size \cite{Newman01}.
We averaged the percolation threshold values obtained 
by these methods to get the final estimate.
Critical exponents $\nu$ and $\beta$ are obtained by the derivative of
percolation probability and $P_{\infty}$, respectively \cite{Lobb80,Martins03}.
The correction-to-scaling \cite{Ballesteros99,Ziff11} is ignored.

Most of the results were produced by using Mersenne twister 
pseudo-random-number generator (MT19937) \cite{MT},
which was confirmed reliable in a site percolation problem \cite{Lee08}.
We also confirmed that other pseudo-random-number generators give equivalent results.

\section{Results}

\begin{figure*}[!tb]
\centering
\includegraphics[width=16cm]{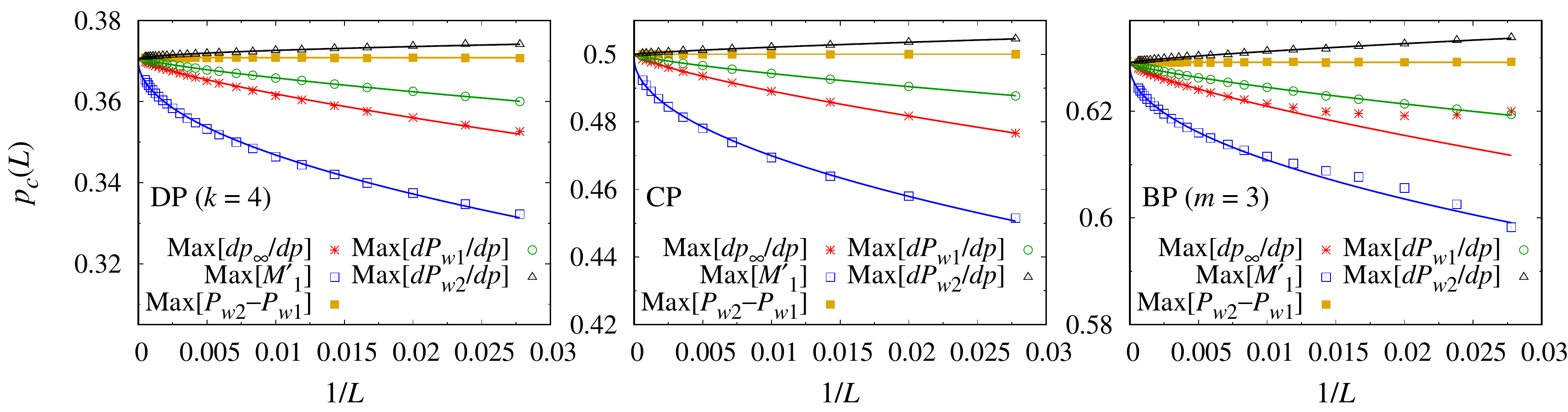}
\caption{The percolation threshold estimate of finite systems $p_c(L)$ 
as a function of linear size $L$ 
for DP ($k=4$), CP, and BP ($m=3$) in the triangular lattice.
They were determined by the probabilities of initial filling ($p$) that give the maximum of 
$dP_{\infty}/dp$, $M'_{1}$, $d(P_{w2}-P_{w1})/dp$, $dP_{w1}/dp$, and $dP_{w2}/dp$.
Solid lines are from fitting of $[p_c(L) - p_c^{(\infty)}] \sim L^{-a}$.
In the right panel, small systems ($L<100$) were excluded in fitting
for $\mathrm{Max}[dP_{\infty}/dp]$ and $\mathrm{Max}[M'_{1}]$.}
\label{pc_vs_L}
\end{figure*}

\begin{figure*}[!tb]
\centering
\includegraphics[width=16cm]{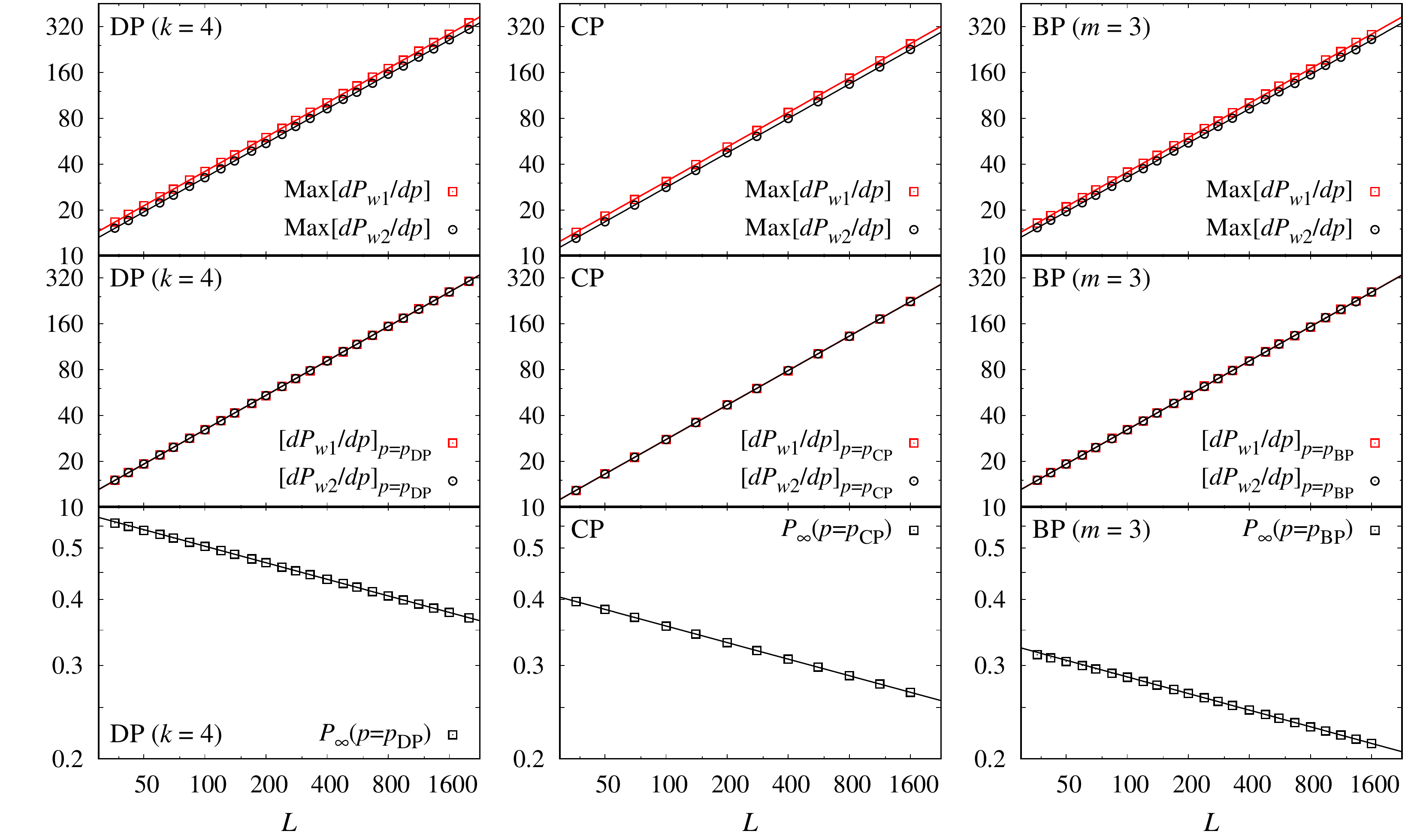}
\caption{Maximum values of $dP_{w1}/dp$ and $dP_{w2}/dp$ in the upper panels, and
values of $dP_{w1}/dp$, $dP_{w2}/dp$, and $P_{\infty}$ at the percolation
threshold $p_c^{(\infty)}$ in the middle and lower panels as a function of
system's linear size $L$ in the triangular lattice in log-log scale. 
Solid lines are from fitting.
In the right lower panel, small systems ($L<100$) were excluded in fitting.}
\label{crit_exp}
\end{figure*}

\begin{table*}[!tb]
\caption{\label{table1} Name, coordination number ($\Delta$), 
percolation threshold ($p_{\mathrm{\scriptscriptstyle BP}}^m$ 
and $p_{\mathrm{\scriptscriptstyle DP}}^k$), and critical exponents ($\nu$ and $\beta$) for the BP and DP 
with the continuous percolation transition of the eleven Archimedean lattices.}
\centering
\begin{tabular}{|cc|cccc|cccc|}
\hline
      &  & \multicolumn{4}{|c|}{Bootstrap percolation} & \multicolumn{4}{|c|}{Diffusion percolation} \\
        \cline{3-10}
Lattice & $\Delta$ & $m$ & $p_{\mathrm{\scriptscriptstyle BP}}^m$  & $\nu$  & $\beta/\nu$ & $k$ & $p_{\mathrm{\scriptscriptstyle DP}}^k$
        & $\nu$  & $\beta/\nu$ \\ \hline
T1          & 6 & $1$ & 0.49997(4) & 1.336(3) & 0.104(1) & $6$ & 0.50000(1) & 1.335(1) & 0.104(1) \\
($3^6$)     &   & $2$ & 0.49999(4) & 1.336(3) & 0.104(1) & $5$ & 0.49999(2) & 1.334(1) & 0.104(1) \\
            &   & $3$ & 0.62915(5) & 1.335(3) & 0.104(2) & $4$ & 0.37083(4) & 1.334(2) & 0.103(1) \\
             \hline
T2          & 4 & $1$ & 0.59272(3) & 1.335(4) & 0.104(1) & $4$ & 0.54731(1) & 1.333(4) & 0.104(1) \\
($4^4$)     &   & $2$ & 0.59272(3) & 1.336(3) & 0.104(1) & $3$ & 0.42037(2) & 1.333(2) & 0.104(1) \\
             \hline
T3          & 3 & $1$ & 0.69710(5) & 1.337(3) & 0.104(1) & $3$ & 0.56008(1) & 1.333(1) & 0.104(1) \\
($6^3$)     &   & $2$ & 0.69703(4) & 1.335(3) & 0.104(1) & $2$ & 0.30943(2) & 1.333(2) & 0.104(1) \\
             \hline
T4          & 5 & $1$ & 0.57948(3) & 1.337(3) & 0.104(1) & $5$ & 0.57950(1) & 1.332(2) & 0.104(1) \\
($3^4,6$)   &   & $2$ & 0.57948(3) & 1.336(3) & 0.104(1) & $4$ & 0.48450(1) & 1.334(1) & 0.104(1) \\
            &   & $3$ & 0.73227(3) & 1.333(2) & 0.103(3) & $3$ & 0.26936(3) & 1.334(1) & 0.104(1) \\
             \hline
T5          & 5 & $1$ & 0.55020(3) & 1.337(3) & 0.104(1) & $5$ & 0.54387(1) & 1.333(1) & 0.104(1) \\
($3^3,4^2$) &   & $2$ & 0.55020(4) & 1.335(3) & 0.104(1) & $4$ & 0.47648(2) & 1.333(2) & 0.104(1) \\
            &   & $3$ & 0.71884(4) & 1.330(4) & 0.103(1) & $3$ & 0.28165(1) & 1.332(1) & 0.103(1) \\
             \hline
T6           & 5& $1$ & 0.55080(3) & 1.337(4) & 0.104(1) & $5$ & 0.54108(1) & 1.333(1) & 0.104(1) \\
($3^2,4,3,4$)&  & $2$ & 0.55080(4) & 1.336(3) & 0.104(1) & $4$ & 0.47072(1) & 1.334(1) & 0.104(1) \\
             &  & $3$ & 0.72813(5) & 1.336(4) & 0.101(4) & $3$ & 0.27194(1) & 1.335(1) & 0.103(1) \\
             \hline
T7          & 4 & $1$ & 0.62180(3) & 1.337(3) & 0.104(1) & $4$ & 0.57502(1) & 1.332(2) & 0.104(1) \\
($3,4,6,4$) &   & $2$ & 0.62178(4) & 1.336(3) & 0.104(1) & $3$ & 0.42652(1) & 1.333(1) & 0.104(1) \\
            &   & $3$ & 0.86713(5) & 1.337(4) & 0.103(3) & $2$ & 0.13447(3) & 1.334(2) & 0.104(1) \\
             \hline
T8          & 4 & $1$ & 0.65268(3) & 1.336(5) & 0.104(1) & $4$ & 0.58661(1) & 1.334(2) & 0.104(1) \\
($3,6,3,6$) &   & $2$ & 0.65269(4) & 1.337(4) & 0.104(1) & $3$ & 0.39451(2) & 1.335(2) & 0.104(1) \\
             \hline
T9          & 3 & $1$ & 0.80787(3) & 1.338(4) & 0.104(1) & $3$ & 0.65335(1) & 1.332(1) & 0.104(1) \\
($3,12^2$)  &   & $2$ & 0.80787(3) & 1.338(3) & 0.104(1) & $2$ & 0.34028(4) & 1.333(2) & 0.104(1) \\
             \hline
T10         & 3 & $1$ & 0.74779(3) & 1.339(4) & 0.104(1) & $3$ & 0.61644(1) & 1.332(1) & 0.104(1) \\
($4,6,12$)  &   & $2$ & 0.74779(3) & 1.337(4) & 0.104(1) & $2$ & 0.31816(3) & 1.333(1) & 0.104(1) \\
             \hline
T11         & 3 & $1$ & 0.72971(3) & 1.336(5) & 0.104(1) & $3$ & 0.58862(1) & 1.334(2) & 0.104(1) \\
($4,8^2$)   &   & $2$ & 0.72971(4) & 1.336(5) & 0.104(1) & $2$ & 0.30280(4) & 1.334(1) & 0.104(1) \\
             \hline
\end{tabular}
\end{table*}

Figure~\ref{P_vs_p} shows the strength of the largest cluster $P_{\infty}(p,L)$,
average cluster size excluding the largest one $M'_{1}(p,L)$, 
percolation probability in any direction $P_{w1}(p,L)$ and in both directions $P_{w2}(p,L)$, 
and proportion of occupied sites $P_{o}(p,L)$ for the DP ($k=4$), CP, and BP ($m=3$) in the triangular lattice.
Parameter $L$ is the linear size of the lattice.
They all show continuous phase transition, which becomes sharper as the lattice size increases.
The proportion of occupied sites $P_{o}(p,L)$ is independent of lattice size and 
does not show any critical behavior at the percolation threshold.
Equivalent behavior is observed in the BP and DP of the other values of $m\leq m_c$ and $k \geq k_c$,
and in the other Archimedean lattices.
The percolation threshold of infinite lattices [$p_c^{(\infty)}=\lim_{L\rightarrow\infty}p_c(L)$]
is determined by the finite-size-scaling: $[p_c(L) - p_c^{(\infty)}] \sim L^{-a}$. 
The percolation threshold estimate for a finite lattice
$p_c(L)$ is determined by the probabilities of
initial filling ($p$) that give the maximum values of $dP_{\infty}(p,L)/dp$, $M'_{1}(p,L)$,
$[P_{w2}(p,L)-P_{w1}(p,L)]$, $dP_{w1}(p,L)/dp$, and $dP_{w2}(p,L)/dp$.
The fitting parameter $a$ depend on lattice structure, 
the physical quantity measured, and percolation type \cite{Ziff02}.
Figure~\ref{pc_vs_L} confirms the scaling behavior very well.
In the case of the BP of $m=3$, however, deviation from the scaling is large in small lattices 
for $dP_{\infty}(p,L)/dp$ and $M'_{1}(p,L)$, 
and so results of lattices smaller than $L=100$ were excluded in the fitting. 
This kind of deviation is also observed in BP of $m=3$ in the other kinds of lattices.
The percolation threshold can also be found by the value of $p$ at the crossing points 
of $P_{w1}(p,L)$ and $P_{w2}(p,L)$ with various linear size ($L$) \cite{Newman01}, 
as shown in Fig.~\ref{P_vs_p}.
We ruled out the data from small lattices to reduce possible error from the finite-size-effect.
All the values of the percolation threshold obtained by these methods were consistent
with each other within error bars. 
The maximum of the probability of percolation 
only in one direction, $P_{w2}(p,L)-P_{w1}(p,L)$,
which has negligible finite-size-effect, gives the most accurate percolation threshold. 

The final estimate of the percolation threshold was obtained by taking an average.
Table~\ref{table1} shows the percolation threshold of all continuous transitions;
cases with the first-order transition (the BP with $m>m_c$ and DP with $k<k_c=\Delta+1-m_c$) 
are omitted in the table.
Note that $m_c=\Delta/2$ for even-coordinated lattices and 
$m_c=(\Delta+1)/2$ for odd-coordinated lattices, 
with exception of T7 (bounce lattice), which has $\Delta=4$ and $m_c=3$.
The percolation threshold results of the BP with $m=1$ or $m=2$ are the same 
as that of the CP within error bars, as is expected.
We confirmed that they are also consistent with exact or the most precise numerical results
of the CP \cite{Sykes64,Suding99,Jacobsen14,Jacobsen15} within relative errors of the order of $0.001\%$.
The percolation threshold values of the BP and DP of other values of $m$ and $k$
for square, triangular, and honeycomb lattices are also
consistent with references \cite{Adler88,Chaves95,Medeiros97},
but our work is much more precise.

Figure~\ref{crit_exp} shows maximum values of the derivative of percolation probability,
$dP_{w1}(p,L)/dp$ and $dP_{w2}(p,L)/dp$, and 
derivative of percolation probability and strength of the largest cluster $P_{\infty}(p,L)$
at the percolation threshold $p_c^{(\infty)}$ calculated in this work.
They satisfy the scaling relations \cite{Lobb80,Martins03}: 
\begin{eqnarray}
\mathrm{Max}\left[dP_{w1}(p,L)/dp\right] \sim L^{1/\nu}, \\
~ \mathrm{Max}\left[dP_{w2}(p,L)/dp\right] \sim L^{1/\nu},\\
\left[dP_{w1}(p,L)/dp\right]_{p=p_c^{(\infty)}} \sim L^{1/\nu}, \\
~ \left[dP_{w2}(p,L)/dp\right]_{p=p_c^{(\infty)}} \sim L^{1/\nu},\\
\mbox{and } \; P_{\infty}(p_c^{(\infty)},L) \sim L^{-\beta/\nu} .
\end{eqnarray}
Therefore, the critical exponents can be obtained by fitting.
In cases of the BP of $m=3$, there exists small but systematic deviation from the
scaling behavior for $P_{\infty}(p_c^{(\infty)},L)$ in small lattices ($L<100$), 
which were excluded from the fitting.
Table~\ref{table1} summarizes the critical exponents obtained in this work.
They are consistent with those of the two-dimensional classical percolation model
($\nu=4/3$ and $\beta/\nu=5/48$) within error bars, 
which are much smaller than references \cite{Kogut81,Branco84,Branco86,Adler88,Adler91,Chaves95,Medeiros97}.
Therefore, we are convinced that continuous transitions of the BP and DP have the same critical exponents
as the CP in two dimensions.

\section{Summary}

We extended the Newman-Ziff algorithm of the classical percolation 
to propose very efficient algorithms for the BP and DP models. 
Using these algorithms we studied the BP and DP in the eleven Archimedean lattices.
The BP with $m\leq m_c$ and the DP with $k\geq (\Delta+1-m_c)$ have
continuous percolation phase transitions.
We found that $m_c=\lfloor (\Delta+1)/2 \rfloor$ except for the bounce lattice (T7), 
which has $m_c=\lfloor (\Delta+1)/2 \rfloor+1$.
Through the finite-size-scaling, 
we calculated the percolation threshold and critical exponents 
for the BP and DP with the continuous phase transition.
We found that the critical exponents $\nu$ and $\beta$ are the same as those of the CP within error bars
to conclude that the BP and DP almost certainly 
belong to the same universality class as the CP in two dimensions.

The algorithms presented in this paper can be directly applied to any dimensions and graphs. Since the BP and DP models are useful both in materials on lattices and in complex systems, studies of the BP and DP using these new algorithms in three-dimensional lattices and complex networks would be also interesting.

\section*{Acknowledgments}
This work was supported by GIST Research Institute (GRI) grant funded by the GIST in 2019.

\bibliographystyle{elsarticle-num}
\bibliography{perc}

\end{document}